\documentclass[12pt]{article}

\usepackage[comma,square,sort&compress]{natbib}
\usepackage{epsfig}
\usepackage{graphicx}

\newcommand{\AHEP}{ \small Instituto de F\'{\i}sica Corpuscular --
  C.S.I.C./Universitat de Val{\`e}ncia \\
  \small Edificio Institutos de Paterna, Apt 22085,
  E--46071 Valencia, Spain\\}
\DeclareMathAlphabet{\mathsc}{OT1}{cmr}{m}{sc}
\newcommand{\TM} {\texttt{TM}}
\newcommand{\TMs} {\texttt{TM}s}
\newcommand{\CL} {\% C.L.}

\newcommand{\Sol}  {\mathsc{sol}}
\newcommand{\Atm}  {\mathsc{atm}}

\newcommand{\Dms}{\Delta m^2_\Sol}
\newcommand{\Dma}{\Delta m^2_\Atm}

\let\vev\VEV
\newcommand{\lesssim}{\:\mbox{\raisebox{-3pt}{$\stackrel%
{\displaystyle <}{\sim}$}}\:}

\newcommand{\la}{|\mathbf{\Lambda}|}

\begin{document}

\title{
\normalsize \hfill IFIC/02-38 \\
\normalsize \hfill UWThPh-2002-25\\[.5cm] 
\large \bf
Constraining Majorana neutrino electromagnetic properties from the
LMA-MSW solution of the solar neutrino problem}
\author{W. Grimus$^a$, M. Maltoni$^b$, T. Schwetz$^a$, \\
  M.A. T\'ortola$^b$ and J.W.F. Valle$^b$ \\[2mm]
  \small $^a$ Institut f\"ur Theoretische Physik, Universit\"at Wien \\
  \small Boltzmanngasse 5, A--1090 Wien, Austria \\[1mm]
  \small $^b$\AHEP}

\date{August 13, 2002}

\maketitle
\begin{abstract}
  In this paper we use solar neutrino data to derive stringent bounds
  on Majorana neutrino transition moments (\TMs). Should such be
  present, they would contribute to the neutrino--electron scattering
  cross section and hence alter the signal observed in
  Super-Kamiokande.  Motivated by the growing robustness of the
  LMA-MSW solution of the solar neutrino problem indicated by recent
  data, and also by the prospects of its possible confirmation at
  KamLAND, we assume the validity of this solution, and we constrain
  neutrino \TMs\ by using the latest global solar neutrino data. We
  find that all elements of the \TM\ matrix can be bounded at the same
  time. Furthermore, we show how reactor data play a complementary
  role to the solar neutrino data, and use the combination of both
  data sets to
  improve the current bounds. Performing a simultaneous fit of LMA-MSW
  oscillation parameters and \TMs\ we find that $6.3 \times
  10^{-10}\mu_B$ and $2.0 \times 10^{-10}\mu_B$ are the 90\% C.L.\ 
  bounds from solar and combined solar + reactor data, respectively.
  Finally, we perform a simulation of the upcoming Borexino experiment
  and show that it will improve the bounds from today's data by
  roughly one order of magnitude.
\end{abstract}

\newpage

\section{Introduction}

Present
solar~\cite{Cleveland:nv,Abdurashitov:2002nt,GaGNO,Fukuda:2002pe,SNONC,SNODN}
and atmospheric neutrino data \cite{atm02,Fukuda:1998mi} provide the
first robust evidence for neutrino flavour conversion and, consequently,
the first solid indication for new physics. 
Neutrino oscillations constitute the most popular explanation for the
data (for recent analyses see
Refs.~\cite{Maltoni:2002ni,solfits,3flavors})
and are a natural outcome of gauge theories of neutrino
mass~\cite{Schechter:1980gr,rev}.
Non-zero neutrino masses manifest themselves through non-standard
neutrino electromagnetic properties.
If the lepton sector in the Standard Model (SM) were minimally
extended in analogy with the quark sector, neutrinos get Dirac masses
($m_\nu$) and their magnetic moments (MMs) are tiny \cite{Fujikawa:yx},
\begin{equation}
  \label{eq:dirac}
\mu_\nu \simeq 3 \times 10^{-19}
\mu_B \left( \frac{m_\nu}{1 \, \mathrm{eV}} \right) \,,
\end{equation}
where $\mu_B$ is the Bohr magneton. Laboratory experiments give 
90\CL\ bounds on the neutrino MMs of $1.8 \times 10^{-10} \mu_B$
\cite{Derbin:wy,Derbin:ua} and $1.3 \times 10^{-10} \mu_B$ \cite{munu}
for the electron neutrino, $6.8 \times 10^{-10} \mu_B$ for the muon
neutrino \cite{LSND} and $3.9 \times 10^{-7} \mu_B$ for the tau
neutrino \cite{DONUT} (see also Ref.~\cite{Hagiwara:pw}). On the other
hand, astrophysics and cosmology provide limits of the order of
$10^{-12}$ to $10^{-11}$ Bohr magnetons \cite{Raffelt:gv}. Improved
sensitivity for the electron neutrino MM from reactor neutrino experiments
is expected, while a tritium $\bar\nu_e$ source experiment \cite{H3}
aims to reach $3 \times 10^{-12} \mu_B$.

It has for a long time been noticed, on quite general ``naturality''
grounds, that Majorana neutrinos constitute the typical outcome of
gauge theories~\cite{Schechter:1980gr}. On the other hand, precisely
such neutrinos also emerge in specific classes of unified theories, in
particular, in those employing the seesaw mechanism~\cite{rev}.  If
neutrinos are indeed Majorana particles the structure of their
electromagnetic properties differs crucially from that of Dirac
neutrinos~\cite{Schechter:1981hw}, being characterized by a $3 \times
3$ complex antisymmetric matrix $\lambda$, the so-called Majorana
transition moment (\TM) matrix. It contains MMs as well as electric
dipole moments of the neutrinos.
The existence of any electromagnetic neutrino moment well above the
expectation in Eq.~(\ref{eq:dirac}) would signal that some special
mechanism---which goes beyond the SM---is at work.  Therefore,
neutrino electromagnetic properties are interesting probes of new
physics.

As noted in Ref.~\cite{Schechter:1981hw} the existence of Majorana
transition moments leads to the phenomenon of spin-flavour precession,
when they move through a magnetic field, as might happen in the Sun.
This was subsequently shown to affect the propagation of solar
neutrinos in an important way, due to the effects of
matter~\cite{Akhmedov:uk}, and this has recently been shown to provide
excellent fits to current data~\cite{Miranda}.

In the present paper we will take a different attitude.  Motivated by
the now robust status of the LMA-MSW solution of the solar neutrino
problem indicated by recent data~\cite{Maltoni:2002ni,solfits}, and
also by the prospects of its possible confirmation by the KamLAND
experiment~\cite{kamland}, we take up this solution as the basis of our
investigation. It has been shown in Ref.~\cite{Miranda} that in this
case the effect of a solar magnetic field can be safely neglected.
Therefore, our present approach is justified and complementary to that
in Ref.~\cite{Miranda}: there we were concerned with
spin-flavour-conversions of solar neutrinos, sensitive to the product
of $\lambda$ times a solar magnetic field, while here we are concerned
with the effect of \TMs\ in the neutrino detection process. Sizable
\TMs\ would contribute to the elastic neutrino--electron scattering in
the Super-Kamiokande (SK) experiment~\cite{Fukuda:2002pe}, and hence,
data from this experiment can be used to constrain electromagnetic
neutrino properties~\cite{Beacom:1999wx,joshipura}.

Here we will tacitly employ the results of the latest analysis in
Ref.~\cite{Maltoni:2002ni} as well as the fit strategies used there,
to perform a fit to the global solar neutrino data in terms of LMA-MSW
oscillation parameters and neutrino \TMs, in order to derive bounds on
the latter. It is important to clarify for which combinations of the
three complex moments the bounds are valid. We will show that all
three moments can be bounded, however, not all at the same level. In
addition, we will show that data from reactor neutrino
experiments~\cite{Derbin:wy,Derbin:ua,munu} which use elastic
neutrino--electron scattering for neutrino detection are to some extent
complementary to the solar neutrino data, so that by combining both
data sets one can improve the bounds.  We also explore the potential
of the upcoming Borexino experiment~\cite{borexino,Alimonti:2002xc} in
probing the \TMs, showing that it is substantially more 
sensitive to the \TMs\ than present experiments.

The paper is organized as follows. In Section~\ref{sec:theor-fram} we
give our theoretical framework and we discuss in detail which elements
of the \TM\ matrix are constrained. In Section~\ref{sec:stat} we present
the statistical method we apply to calculate bounds on the \TMs, and
in Section~\ref{sec:analys-solar-react} we describe our analysis of solar
and reactor neutrino data. The resulting bounds are discussed in
Section~\ref{sec:bounds-la-from} and the sensitivity of the upcoming
Borexino experiment is considered in
Section~\ref{sec:sensitivity-borexino}. A brief discussion concludes the
paper in Section~\ref{sec:conclusions}.

\section{The theoretical framework}
\label{sec:theor-fram}

In this paper,
we adopt the following basic working assumptions:
\begin{itemize}
\item Only three light neutrinos exist, doublets under $SU(2) \otimes
  U(1) $ (no light sterile neutrinos); this is well-motivated by
  recent global fits of all available neutrino oscillation data, including
  also those that follow from short baseline
  searches~\cite{Maltoni:2002xd}.
\item The solution of the solar neutrino puzzle is given by the
  LMA-MSW solution, as indicated by recent global fits of solar
  data~\cite{solfits,3flavors}.
\item Neutrinos have Majorana nature as expected from
  theory~\cite{rev,Schechter:1981hw}.
\item Neutrinos are endowed with \TMs\ arising from some
  unspecified electroweak gauge theory.
\end{itemize}
Although the latter is probably the least motivated of our
assumptions, there are some models of the electroweak interaction
where the possibility of enhanced \TMs\ has been discussed~\cite{rev}.

In experiments where the neutrino detection reaction is elastic
neutrino--electron scattering, like in SK, Borexino and some reactor
experiments~\cite{Derbin:wy,Derbin:ua,munu}, the electromagnetic cross
section is~\cite{Bardin:wr}
\begin{equation}\label{cross}
\frac{d\sigma_\mathrm{em}}{dT} = \frac{\alpha^2 \pi}{m_e^2 \mu_B^2}
\left(\frac{1}{T} - \frac{1}{E_\nu}\right) \mu^2_\mathrm{eff} \,,
\end{equation}
where the effective MM square is given by~\cite{GS}
\begin{equation}\label{mmeff}
\mu^2_\mathrm{eff} =  
a_-^\dagger \lambda^\dagger \lambda a_- +
a_+^\dagger \lambda \lambda^\dagger a_+
\,.
\end{equation}
The electromagnetic cross section adds to the weak cross section and allows
to extract information on the \TM\ matrix $\lambda$. In this cross section,
$T$ denotes the kinetic energy of the recoil electron, $E_\nu$ is the energy
of the incoming neutrino and the 3-vectors $a_-$ and $a_+$ denote the
neutrino amplitudes for negative and positive helicities, respectively, at
the detector. The square of the effective MM given in Eq.~(\ref{mmeff}) is
independent of the basis chosen for the neutrino state \cite{GS}. In what
follows we consider both the flavour basis and the mass eigenstate basis. We
will use the convention that $a_\mp$ and $\lambda = (\lambda_{\alpha\beta})$
denote the quantities in the flavour basis, whereas in the mass basis we
will use $\tilde a_\mp$ and $\tilde\lambda =(\lambda_{jk})$.  The two bases
are connected via the neutrino mixing matrix $U$:
\begin{equation}\label{trafo}
\tilde a_- = U^\dagger a_- \,, \quad
\tilde a_+ = U^T       a_+ \,, \quad
\tilde \lambda = U^T \lambda U \,.
\end{equation}
One usually decomposes the transition moment matrix as
\begin{equation}\label{decomp}
\lambda = \mu - i d \,,
\end{equation}
where $\mu$ and $d$ are hermitian matrices. On general grounds these
matrices are, in addition, antisymmetric and imaginary for Majorana
neutrinos~\cite{Schechter:1981hw} (see also Ref.~\cite{GS} and
references cited therein). It is, furthermore, useful to
define vectors $\mathbf\Lambda = (\Lambda_\alpha)$ and
$\tilde{\mathbf\Lambda} = (\Lambda_j)$ in the flavour and mass basis,
respectively, by
\begin{equation}\label{defL}
\lambda_{\alpha\beta} = \varepsilon_{\alpha\beta\gamma} \Lambda_\gamma
\quad \mathrm{and} \quad 
\lambda_{jk} = \varepsilon_{jkl} \Lambda_l \,.
\end{equation}
Thus, in the flavour basis we have
$\lambda_{e \mu} = \Lambda_\tau$,
$\lambda_{\mu\tau} = \Lambda_e$ and
$\lambda_{\tau e} = \Lambda_\mu$.
Note also that 
\begin{equation}
\label{basisindep}
\la^2 = \frac{1}{2}\, \mathrm{Tr} \left( \lambda^\dagger \lambda \right)
\quad \Rightarrow \quad
\la = |\tilde{\mathbf\Lambda}| \,.
\end{equation}
This means if we are able to find a bound on $\la$, we have not only 
constrained the \TMs\ in the flavour basis but also in the 
mass basis. In the numerical section of the paper, this will be 
exactly our strategy.

Let us now discuss the detailed form that the effective MM 
square $\mu_\mathrm{eff}^2$ in Eq.~(\ref{mmeff}) takes when we
assume the LMA-MSW solution of the solar neutrino problem, denoted by
$\mu_\mathrm{LMA}^2$.
Numerical integrations of the neutrino evolution equations performed
in Ref.~\cite{Miranda} showed that for an effective MM of
$10^{-11}\mu_B$ a solar magnetic field of the order of 80 KGauss has
practically no effect on the LMA-MSW solution. The results of
Ref.~\cite{Miranda} imply that it is safe to neglect possible solar
magnetic field effects on the evolution of the neutrino state in the
sun in case of LMA-MSW. This follows essentially from the fact that
the terms in the evolution Hamiltonian related to the LMA-MSW
mass-squared difference and the matter potential are much bigger than
the electromagnetic terms. Hence, helicity is conserved in solar
neutrino propagation, so that we can set $a_+ = 0$, thereby
eliminating the second term in Eq.~(\ref{mmeff}).  Another good
approximation used in fitting solar data in a three-neutrino
scenario~\cite{3flavors} is to set $\Dma \gg \Dms
\equiv \Delta m^2$, \textit{i.e.},
oscillations with $\Dma$ are averaged out. Using the parameterization
(\ref{defL}) we obtain from Eq.~({\ref{mmeff})
\begin{equation} \label{eq:mu2effLMA}
   \mu^2_\mathrm{LMA} = \la^2 - P^{3\nu}_{e3} |\Lambda_3|^2 -
\sum_{j,k=1}^2 \vev{ (\tilde{a}^j_-) (\tilde{a}^k_-)^* }
\Lambda_j^* \Lambda_k  \,.
\end{equation}
The brackets $\vev{...}$ in the last term in Eq.~(\ref{eq:mu2effLMA})
denote the average over the production point, earth-sun distance and
zenith angle, and we have made use of the relation 
\begin{equation}\label{P}
P^{3\nu}_{ej} \equiv \vev{ |\tilde{a}_-^j|^2 } \quad (j=1,2,3) 
\end{equation}
for the probability that the neutrino produced in the core of the sun as
a $\nu_e$ arrives at the detector as a mass eigenstate $\nu_j$.
From Eq.~(\ref{eq:mu2effLMA}) one learns that besides the absolute values
of the \TMs\ $|\Lambda_1|,|\Lambda_2|,|\Lambda_3|$ only one complex phase 
$\delta=\mathrm{arg}(\Lambda_1^* \Lambda_2)$ enters into $\mu^2_\mathrm{LMA}$.

Eq.~(\ref{eq:mu2effLMA}) can be further simplified by making use of
the large mass-squared difference $\Delta m^2 \sim 5\times 10^{-5}$
eV$^2$ implied by the LMA-MSW solution. First, because of vacuum
oscillations on the way from the sun to the earth the neutrino state
arriving at the earth can be considered as an incoherent mixture of
mass eigenstates~\cite{GS,dighe}. Note also that for the LMA-MSW
solution, earth matter effects are very small~\cite{earthmatter} so
that we neglect them in our expression for
$\mu^2_\mathrm{LMA}$.\footnote{Note, however, that the neglection of
  earth matter effects is not a very good approximation
  when performing a combined fit of \TMs\ and
  oscillation parameters. Therefore, in this case we do take into
  account earth 
  matter effects in the minimization of the $\chi^2$ with respect to
  the solar oscillation parameters.} Then we find that only the
diagonal elements of the matrix $\left( {\tilde \lambda}^\dagger
  \tilde \lambda \right)$ contribute, and as a result only the
probabilities (\ref{P}) appear in determining the detected neutrino
signal. Setting $s_{13} \equiv \sin \theta_{13}$, $c_{13} \equiv \cos
\theta_{13}$ and $U_{e3} = s_{13}$ one has, to a very good
approximation~\cite{3flavors},
\begin{equation}\label{P2}
P^{3\nu}_{e3} = s_{13}^2 \quad \mathrm{and} \quad
P^{3\nu}_{ej} = c_{13}^2 P^{2\nu}_{ej} \; (j=1,2)
\quad \mathrm{with} \quad P^{2\nu}_{e1} + P^{2\nu}_{e2} = 1 \,.
\end{equation}
The probabilities $P^{2\nu}_{ej}$ $(j=1,2)$ are the effective
2-neutrino probabilities for the solar neutrino problem, where all the
averages mentioned above have been taken into account.  
Using now the probabilities in Eq.~(\ref{P2}) and the approximations mentioned
above, Eq.~(\ref{eq:mu2effLMA}) becomes
\begin{equation}\label{muLMA}
\mu^2_\mathrm{LMA} = 
|\Lambda_1|^2 + |\Lambda_2|^2 + c_{13}^2 |\Lambda_3|^2 -
c_{13}^2 \sum_{j=1,2} P^{2\nu}_{ej} |\Lambda_j|^2 \,.
\end{equation}

It is well known from reactor neutrino data~\cite{CHOOZ} that the
mixing angle $\theta_{13}$ is rather small.  Fits to the CHOOZ data
and atmospheric neutrino data~\cite{3flavors} show that at 3$\sigma$
one has $s_{13}^2 \lesssim 0.05$. This allows us to replace $c_{13}^2$
in Eq.~(\ref{muLMA}) by 1. Then we arrive at the final formula
\begin{equation}\label{finalLMA}
\mu^2_\mathrm{LMA} = \la^2 - |\Lambda_2|^2 +
P^{2\nu}_{e1} \left( |\Lambda_2|^2 - |\Lambda_1|^2 \right) \,,
\end{equation}
which will be used in the numerical section evaluating solar neutrino
data.  The probability $P^{2\nu}_{e1}$ is a function of the ratio
$\Delta m^2/E_\nu$ and the solar mixing angle $\theta \equiv
\theta_{12}$. In the following we will drop the super-script 
$2\nu$ and $P_{e1}$ or $P_{ee}$ refers always to a 2-neutrino probability.
Eq.~(\ref{finalLMA}) naturally makes no distinction
between MMs and electric dipole moments.  Since we aim at
constraining $\la$, \texttt{all} elements of the \TM\ matrix
will be bounded at the same time.

Now we move to the case of reactor neutrinos. There we have a
pure $\bar\nu_e$ source. Therefore, we have $a_- = 0$ and $a_+ =
(1,0,0)^T$ since in these experiments the baseline is much too short
for any oscillations to develop. The resulting $\mu_\mathrm{eff}^2$
relevant in reactor experiments is given as
\begin{equation}\label{reactorMM}
\mu_\mathrm{R}^2 = |\Lambda_\mu|^2 + |\Lambda_\tau|^2 \,.
\end{equation}
From this relation it is clear that reactor data on its own
\texttt{cannot} constrain all \TMs\ contained in $\lambda$, since
$\Lambda_e$ does not enter in Eq.~(\ref{reactorMM}). In order to combine
reactor and solar data it is useful to rewrite $\mu_\mathrm{R}^2$ in
terms of the mass basis quantities.  With Eq.~(\ref{defL}) we readily
derive
\begin{equation}\label{LLjk}
\left( \tilde \lambda {\tilde \lambda}^\dagger \right)_{jk} =
\la^2 \delta_{jk} - \Lambda_j \Lambda_k^* \,.
\end{equation}
Then with Eq.~(\ref{trafo}) we obtain
\begin{equation} \label{finalreactor}
\mu_\mathrm{R}^2 = 
\la^2 - c^2 |\Lambda_1|^2 - s^2 |\Lambda_2|^2 
-2 s c |\Lambda_1| |\Lambda_2| \cos\delta \,,
\end{equation}
where $c=\cos\theta$ and $s=\sin\theta$, $\theta$ being the solar
mixing angle. Further, we notice that the relative phase
$\delta=\mathrm{arg}(\Lambda_1^* \Lambda_2)$ between $\Lambda_1$ and
$\Lambda_2$ appears in addition to $\la$, $|\Lambda_1|$ and
$|\Lambda_2|$.

\section{Statistical method and qualitative discussion}
\label{sec:stat}

In the following we will use data from solar and reactor neutrino
experiments to constrain neutrino \TMs. The $\chi^2$-function obtained
from the data depends on the solar oscillation parameters
$\tan^2\theta$ and $\Delta m^2$ as well as on the elements of the \TM\ 
matrix $\lambda$.  Regarding the dependence on the oscillation
parameters we will take two different attitudes. One is to assume that
$\tan^2\theta$ and $\Delta m^2$ will be determined with good accuracy
at the KamLAND experiment, and hence, we will consider the $\chi^2$ at
\texttt{fixed} points in the $\tan^2\theta - \Delta m^2$ plane (method
I). In the second approach we will derive a bound on the \TMs\ by
minimizing the $\chi^2$ with respect to $\tan^2\theta$ and $\Delta m^2$
(method II). This second procedure takes into account the present
uncertainty of our knowledge of the oscillation parameters.

Let us describe in detail how we calculate a bound on the \TMs.
Our aim is to constrain $\la$, therefore it is convenient to consider
the $\chi^2$ as a function of the independent parameters
$\la,|\Lambda_1|,|\Lambda_2|$ and $\delta$. As discussed in the
previous section, $\delta$ appears only if reactor data are included.
The $\chi^2$-functions which we are using for the individual data sets
(solar rates, SK recoil electron spectrum, reactor data) will be
described in detail in the following sections.  When performing 
the fit to the data, we find that in general the minimum of the 
$\chi^2$ occurs close or outside the physical boundary of the 
parameters $\la,|\Lambda_1|,|\Lambda_2|$. To take this into account we apply
Bayesian techniques to calculate an upper bound on $\la$. 
We minimize the $\chi^2$ with respect to
$|\Lambda_1|$, $|\Lambda_2|$ and $\delta$ for each value of $\la$, taking
into account the allowed region $0 \le |\Lambda_1|^2 + |\Lambda_2|^2 \le \la^2$:
\begin{equation}\label{minL}
\chi^2(\la)= \mathrm{Min}  \left[
\chi^2(\la,|\Lambda_1|,|\Lambda_2|,\delta) \right] \,.
\end{equation}
In method I we do this for fixed values of the oscillation parameters,
scanning over the LMA-MSW region, whereas in method II we minimize
also with respect to $\tan^2\theta$ and $\Delta m^2$ in
Eq.~(\ref{minL}). Then the $\chi^2$ is transformed into a likelihood
function via
\begin{equation}
\mathcal{L} \propto \exp\left( -\frac{1}{2} \chi^2 \right) \,.
\end{equation}
Now we use Bayes' theorem and a flat prior distribution $p(\la)$ in the
physically allowed region, $p(\la)=\Theta(\la)$, to obtain a
probability distribution for $\la$:
\begin{equation}
f(\la) \, d\la= \frac{\mathcal{L}(\la) \, \Theta(\la) \, d\la }
{\int_0^\infty \mathcal{L}(\la') \, d\la' } \,.
\end{equation}
An upper bound $b_\alpha$ on $\la$ at a C.L.\ $\alpha$ is given by the
equation
\begin{equation}
\int_0^{b_\alpha} f(\la)d\la = \alpha \,.
\end{equation}

Let us consider in more detail the minimization with respect to
$|\Lambda_1|$ and $|\Lambda_2|$ in Eq.~(\ref{minL}) in the case of
solar data without reactor. As we will show later there is no evidence
for a non-zero $\mu_\mathrm{LMA}^2$ in the data. Therefore the minimum
of the $\chi^2$ for a given $\la$ will occur if $\mu_\mathrm{LMA}^2$
is minimal.  Departing from Eq.~(\ref{finalLMA}) it is easy to show
that
\begin{equation}\label{mumin}
\mathrm{Min}\, [ \mu^2_\mathrm{LMA} ] = \left\{
\begin{array}{l@{\quad}l}
\la^2 \, P_{e1} & \mathrm{for}\quad P_{e1}\le 0.5 \,, \\
\la^2 \, (1-P_{e1}) & \mathrm{for}\quad P_{e1} > 0.5 \,.
\end{array} \right.
\end{equation}
The minimum occurs at
\begin{equation}\label{mumin2}
\renewcommand{\arraystretch}{1.2}
\la^2 = \left\{
\begin{array}{l@{\quad\mbox{for}\quad}l}
 |\Lambda_2|^2  &  P_{e1} < 0.5 \,,\\
|\Lambda_1|^2 + |\Lambda_2|^2
&  P_{e1} = 0.5 \,,\\
 |\Lambda_1|^2 & P_{e1} > 0.5 \,.
\end{array}
\right.
\end{equation}
From Eq.~(\ref{mumin}) follows that the bound on $\la$ is strongest if
$P_{e1}=0.5$ because in this situation $\mu_\mathrm{LMA}^2$ is
maximal. In Fig.~\ref{fig:Pe1} we show contours of constant $P_{e1}$
in the $\tan^2\theta - \Delta m^2/E_\nu$ plane.  For definiteness, the
probabilities in the figure are obtained by performing the averaging
over the production distribution inside the sun for the $^7$Be flux
most relevant for Borexino. However, the probabilities for the $^8$B
flux relevant for SK are very similar.
\begin{figure}[t] \centering 
\includegraphics[width=0.7\linewidth]{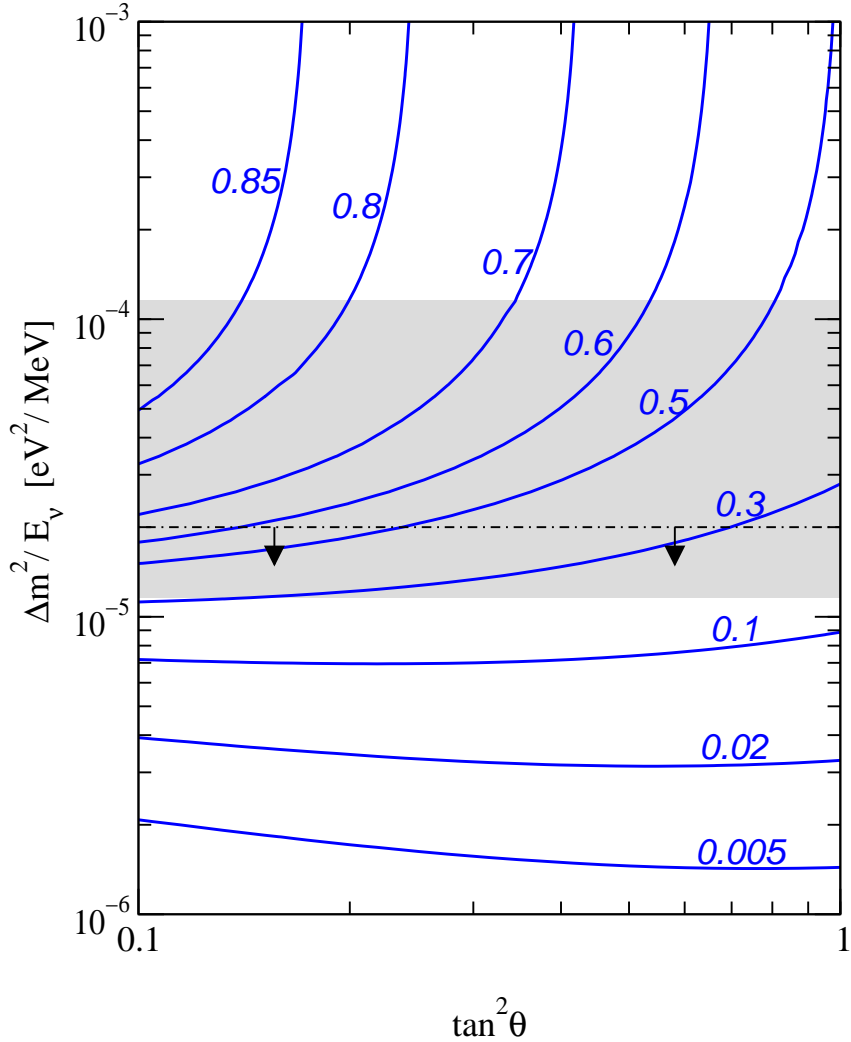} 
\caption{Contours of constant $P_{e1}$. 
The shaded region is relevant for Borexino, whereas the region below the
dash-dotted line is relevant for SK.}\label{fig:Pe1}
\end{figure}

The shaded region in Fig.~\ref{fig:Pe1} is the region relevant for
Borexino, assuming a mass-squared difference in the range $10^{-5}$
eV$^2 < \Delta m^2 < 10^{-4}$ eV$^2$, whereas for SK the region below
the dash-dotted line is most important, due to the higher energy of
the $^8$B neutrinos. We can read off from the figure that in most part
of the SK region $P_{e1}$ is very small. According to
Eq.~(\ref{mumin2}) this means that the sensitivity of SK for $\la$ is
limited because of the small $P_{e1}$. In contrast, we expect a much
better sensitivity of Borexino, because, in a large part of the
relevant parameter space in this case, $P_{e1}$ is close to the optimal
value of $0.5$.

The arguments presented here are valid as long as the $\chi^2$ depends
only on the parameter combination given in Eq.~(\ref{finalLMA}). The data from
reactor experiments depend on a different combination of \TM\ parameters
(see Eq.~(\ref{finalreactor})) implying that the combined analysis of
SK and reactor data will potentially lead to more stringent limits
than the SK data alone.

% BEGIN

\section{Analysis of solar and reactor neutrino data}
\label{sec:analys-solar-react}

In this section we describe briefly our analysis of solar neutrino
data and the data which we are using from reactor neutrino
experiments. Neglecting the electromagnetic contribution to the small elastic
scattering signal in SNO, one notices that the neutrino \TMs\
contribute directly to the observed signal only in SK. However,
the uncertainty in the $^8$B flux leads to correlations between
SK and all other experiments. Therefore, even for fixed oscillation
parameters (method I), also other experiments
give some information on \TMs\ and it is important to include them in
the analysis~\cite{joshipura}.

We divide the solar neutrino data into the total rates observed in all
experiments and the SK recoil electron spectrum (with free overall
normalization). Before combining rates with SK spectra, we derive
bounds on $\la$ for each of these cases separately. Subsequently we
consider the reactor data by themselves and in combination with the
global sample of solar data.

\subsection{Solar rates}

In this analysis we use the event rates measured in the Chlorine
experiment Homestake \cite{Cleveland:nv} (Cl), the Gallium experiments
Sage \cite{Abdurashitov:2002nt}, Gallex and GNO
\cite{GaGNO} (GaGNO), the total event rate of
SK \cite{Fukuda:2002pe} based on the 1496 days data
sample and the recent result of the Sudbury neutrino observatory
\cite{SNONC} for the charged current (SNO CC) and neutral
current (SNO NC) solar neutrino rates.
For the analysis we use the $\chi^2$-function
\begin{equation}\label{ratechi2}
\chi^2_\mathrm{rates} = \sum_{j,k} 
(R_j - D_j) (V^\mathrm{rate}_{jk})^{-1} (R_k - D_k) \,.
\end{equation}
Here the indices $j,k$ run over the 6 solar neutrino rates (Cl, Sage, GaGNO,
SK, SNO CC, SNO NC), $D_j$ are the experimental rates, $R_j$ are the
theoretical predictions and $V^\mathrm{rate}_{jk}$ is the covariance
matrix.
\begin{table}
\begin{center} 
\begin{tabular}{|c|c|}
\hline
Cl \cite{Cleveland:nv} & $2.56\pm 0.16\pm 0.16$ SNU\\
Sage \cite{Abdurashitov:2002nt} & 
$70.8\,{^{+5.3}_{-5.2}}\,{^{+3.7}_{-3.2}}$ SNU\\
GaGNO \cite{GaGNO} & $73.3\pm 4.7\pm 4.0$ SNU \\
SK \cite{Fukuda:2002pe} & $0.465 \pm 0.005\,{^{+0.014}_{-0.012}}$ \\
SNO CC \cite{SNONC} & $0.3485 \pm 0.012 \pm 0.018$ \\
SNO NC \cite{SNONC} & $1.008 \pm 0.087 \pm 0.097$ \\
\hline
\end{tabular}
\end{center}
\caption{Solar neutrino rates used in our analysis.
For SK and SNO the given number is the ratio of the observed
flux and the $^8$B flux predicted by the BP00 SSM~\cite{BP00}.}
\label{data}
\end{table}

The data we are using are given in Table \ref{data}. The rates of SK
and SNO are the ratio of the measured flux and the $^8$B flux
predicted by the standard solar model (SSM) BP00~\cite{BP00} and no
disappearance of solar $\nu_e$ ($\Phi^0_\mathrm{B} = 5.05\times 10^6$
cm$^{-2}$s$^{-1}$). Note that the SNO CC and NC fluxes as given in
Ref.~\cite{SNONC} have been obtained from a fit to the SNO
spectral data assuming no distortion of the neutrino energy spectrum,
\textit{i.e.}, a survival probability constant in energy, which in general
is not realized in the case of neutrino oscillations. However, in the
case of the LMA-MSW solution it is a very good approximation to use the
fluxes given in Ref.~\cite{SNONC}.

The theoretical predictions $R_j$ which appear in Eq.~(\ref{ratechi2})
are calculated as described in Ref.~\cite{Maltoni:2002ni}. However, the rate
$R_\mathrm{SK}$ of the SK experiment includes an extra contribution
from the electromagnetic scattering. This rate is given by
\begin{equation}
\label{SKrate}
    R_\mathrm{SK}= \frac{ \int_0^\infty dE_\nu\, \phi_\mathrm{B}(E_\nu)
    \left\{ P_{ee}(E_\nu) \sigma^W_e (E_\nu) +
    [1- P_{ee}(E_\nu)] \sigma^W_\mu (E_\nu) +
    \mu^2_\mathrm{LMA}(E_\nu) \sigma'_\mathrm{em}(E_\nu) \right\}}
    { \int_0^\infty dE_\nu\, \phi_\mathrm{B}(E_\nu) 
    \,\sigma^W_e (E_\nu)} \,,
\end{equation}
where $\phi_\mathrm{B}(E_\nu)$ is the shape of the $^8$B flux
normalized to 1, which we take from Ref.~\cite{bahcallpage}, and
\begin{equation}\label{sigma}
    \sigma(E_\nu)=\int_0^{T_\mathrm{max}(E_\nu)} dT' 
    \frac{d\sigma(T',E_\nu)}{dT'}
    \int_{T_1}^{T_2} dT \, R(T',T) 
    \quad\mbox{with}\quad
    \sigma = \sigma^W_e,\sigma^W_\mu,\sigma'_\mathrm{em} \,.
\end{equation}
We use the SM weak cross sections $\sigma^W_{e,\mu}$
including radiative corrections~\cite{Bahcall:1995mm}, 
$\mu_\mathrm{LMA}^2$ is given in Eq.~(\ref{finalLMA}), and
\begin{equation}\label{sigmaem}
    \frac{d\sigma'_\mathrm{em}(T',E_\nu)}{dT'} =
    \frac{\alpha^2 \pi}{m_e^2\mu_B^2} 
    \left(\frac{1}{T'} - \frac{1}{E_\nu}\right) \,.
\end{equation}
Here $E_\nu$ is the neutrino energy, while $T'$ and $T$ denote the
true and the measured recoil electron kinetic energies, respectively.
The integration in $T$ is from the SK threshold 5 MeV (total
electron energy) up to 20 MeV.  The maximum kinetic energy of the
recoil electrons is given by $T_\mathrm{max}(E_\nu) = 2E_\nu^2/(2E_\nu
+ m_e)$. The resolution function $R(T',T)$ is taken from
Ref.~\cite{Fukuda:1998ua}.

The covariance matrix $V^\mathrm{rate}_{jk}$ in the $\chi^2$ of
Eq.~(\ref{ratechi2}) takes into account the experimental errors and
theoretical uncertainties from detection cross sections
$V_{jk}^\mathrm{CS}$ and SSM predictions $V_{jk}^\mathrm{flux}$ added
in quadrature~\cite{fogli},
\begin{equation}
    V_{jk}^\mathrm{rate} = V_{jk}^\mathrm{exp} + 
    V_{jk}^\mathrm{CS} + V_{jk}^\mathrm{flux} \,.
    \label{Vrates}
\end{equation}
The experimental errors for different experiments are uncorrelated.
For $j,k=$ Cl, Sage, GaGNO, SK we have $V_{jk}^\mathrm{exp} =
\delta_{jk} \sigma^2_j$, where the $\sigma^2_{j}$ are the experimental
uncertainties, calculated by adding in quadrature the statistical and
systematic uncertainties given in
Table~\ref{data}. However, the statistical and systematic errors for
the SNO CC and NC rates are strongly correlated, with
$\rho^\mathrm{stat} = -0.518$ (the same as the correlation for the CC
and NC day-night asymmetry, given in Ref.~\cite{SNODN}) and
$\rho^\mathrm{sys} = -0.508$ (which we derive from Table II of
Ref.~\cite{SNONC}).
The second and third terms in Eq.~(\ref{Vrates}) take into account the
uncertainties in the detection cross sections and in the SSM
predictions of the neutrino fluxes, respectively. 
For details see Refs.~\cite{Maltoni:2002ni,fogli} and references therein.

\subsection{The Super-Kamiokande recoil electron spectrum}

In this section we consider the \texttt{shape} of the SK recoil
electron spectrum. The electromagnetic contribution to the
neutrino--electron scattering cross section
leads to a substantially different spectrum of the
scattered electrons than expected from the SM weak
interaction. Therefore, the SK spectral data provide a useful tool
to constrain neutrino \TMs~\cite{Beacom:1999wx}.

We perform a fit to the latest 1496 days SK data presented in
Ref.~\cite{Fukuda:2002pe}. There the event rates are given in 8 energy
bins ranging from 5 to 20 MeV.  The energy bins 2 to 7 are further
divided into 7 bins of zenith angle, which makes up a total of 44 data
points $D_i$. We define the $\chi^2$-function
\begin{equation}\label{chi2spect}
\chi^2_\mathrm{spect} =
\sum_{i,j=1}^{44} (\alpha R_i - D_i) (V^\mathrm{spect}_{ij})^{-1} 
(\alpha R_j - D_j) \,.
\end{equation}
The theoretical prediction $R_i(\tan^2\theta, \Delta m^2,
\mu^2_\mathrm{LMA})$ for the $i$-th bin is calculated as in
\linebreak
Eq.~(\ref{SKrate}) with the integration interval $[T_1,T_2]$ in
Eq.~(\ref{sigma}) chosen according to the energy interval of the given
bin. 
The covariance matrix $V^\mathrm{spect}_{ij}$ in
Eq.~(\ref{chi2spect}) contains statistical and systematic experimental
errors. These are obtained from Ref.~\cite{Fukuda:2002pe} taking into
account the correct correlations of systematic errors. The $\chi^2$ of
Eq.~(\ref{chi2spect}) is minimized with respect to the normalization
factor $\alpha$ in order to isolate only the shape of the spectrum.

\subsection{Reactor data}

Data from neutrino electron scattering at nuclear reactor
experiments~\cite{Derbin:wy,Derbin:ua,munu} can be used to constrain
the combination of \TMs\ given in Eq.~(\ref{finalreactor}).  Here we
use data from the Rovno nuclear power plant~\cite{Derbin:wy} and from
the Bugey reactor~\cite{munu}.  To include this information in our
analysis we make the following ansatz for the $\chi^2$-function:
\begin{equation}\label{chi2react}
\chi^2_\mathrm{reactor} (\mu_\mathrm{R}) = 
\sum_i
\left( \frac{N^i_\mathrm{weak} + N^i_\mathrm{em}(\mu_\mathrm{R}) - N^i_\mathrm{obs} } 
{\sigma^i} \right)^2 \,.
\end{equation}
The sum is over the two experiments Rovno and Bugey,
$N^i_\mathrm{obs}$ is the observed number of events with the one
standard deviation error $\sigma^i$, $N^i_\mathrm{weak}$ is the number
of events expected in the case of no neutrino \TMs\ (only the standard
weak interaction) and $N^i_\mathrm{em}(\mu_\mathrm{R})$ is the number of events
due to the electromagnetic scattering of neutrinos with an effective
MM $\mu_\mathrm{R}$. We can write the latter as
\begin{equation}\label{Nem}
N^i_\mathrm{em}(\mu_\mathrm{R}) = C^i \,
\left(\frac{\mu_\mathrm{R}}{10^{-10} \mu_B}\right)^2 \,.
\end{equation}
\begin{figure}[t] \centering
\includegraphics[width=0.7\linewidth]{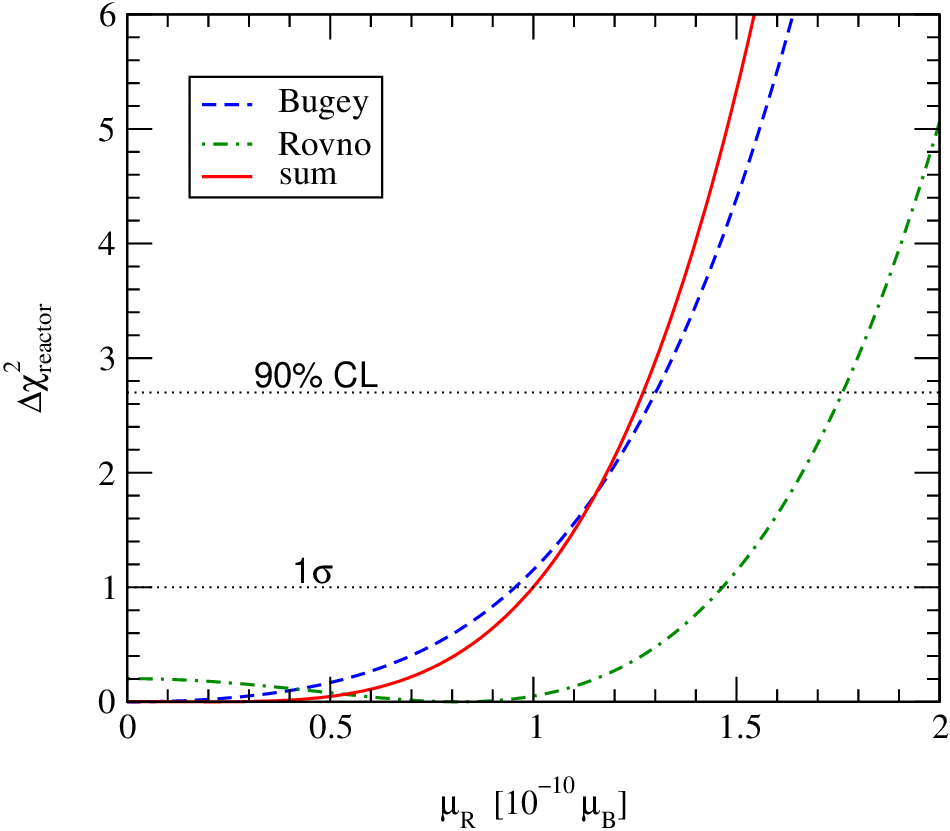}
\caption{$\Delta \chi^2$ of the reactor experiments 
Rovno~\cite{Derbin:wy} and Bugey~\cite{munu} and the total 
$\Delta\chi^2_\mathrm{reactor}$ as a function of the effective
MM.}
\label{fig:react}
\end{figure}

For the Rovno reactor we obtain from Table I of Ref.~\cite{Derbin:wy}
the values $N^\mathrm{Rovno}_\mathrm{obs} = 41$,
$\sigma^\mathrm{Rovno} = 20$ and $N^\mathrm{Rovno}_\mathrm{weak} =
32$.  Furthermore, in this table the number of events due to the
electromagnetic scattering of neutrinos with an effective MM
$\mu_\mathrm{R} = 2 \times 10^{-10} \mu_B$ is given:
$N^\mathrm{Rovno}_\mathrm{em}(2\times 10^{-10} \mu_B)=54$.  Using this
last information the constant in Eq.~(\ref{Nem}) can be determined as
$C^\mathrm{Rovno} = 54/4$.
Results of the MUNU experiment performed at the Bugey reactor have
been presented recently at the Neutrino 2002 conference \cite{munu}.
These lead to the bound $\mu_\mathrm{R} < 1.3\times 10^{-10}\mu_B$ 
at 90 \CL, if
an energy cut of $E_e > 1$ MeV is used.  From Ref.~\cite{munu} we get
(in events per day) $N^\mathrm{Bugey}_\mathrm{obs} = 0.37$,
$\sigma^\mathrm{Bugey} = 0.25$ and $N^\mathrm{Bugey}_\mathrm{weak} =
0.45$.  The constant in Eq.~(\ref{Nem}) can be calculated by making
use of the 90\CL\ bound cited above: we solve the equation
$\chi^2_\mathrm{Bugey}(\mu_\mathrm{R} = 1.3\times 10^{-10}\mu_B) = 2.7$. This
leads to $C^\mathrm{Bugey}=0.196$.  In Fig.~\ref{fig:react} we show
the $\Delta \chi^2$ of the two experiments and also the sum. Note that
with our ansatz Eqs.~(\ref{chi2react}) and (\ref{Nem}) we can
reproduce to good accuracy the 1 $\sigma$ upper bound $\mu_\mathrm{R} <
1.5\times 10^{-10}\mu_B$ from the Rovno data \cite{Derbin:wy}.

\section{Bounds on $\la$ from solar and reactor data}
\label{sec:bounds-la-from}

\begin{figure}[t] \centering
\includegraphics[width=0.7\linewidth]{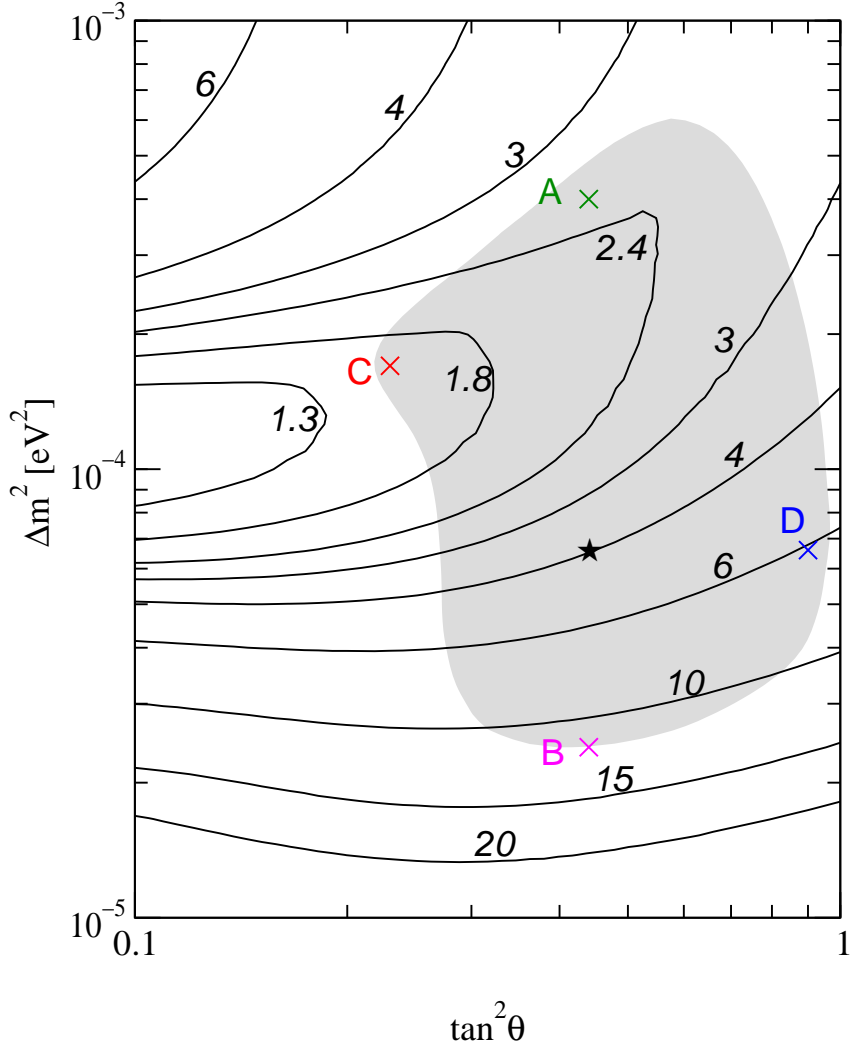}
\caption{Contours of the 90\CL\ bound on $\la$ in units of 
  $10^{-10}\mu_B$ from solar data. The shaded region is the 3$\sigma$
  LMA-MSW region from Ref.~\cite{Maltoni:2002ni}, the best fit point
  is marked with a star. The crosses labeled A, B, C, D are the test
  points considered in Table \ref{tab:bounds}.}
\label{fig:cnt_rat_sp}
\end{figure}
First we discuss the bounds on $\la$ from solar data alone.  Fixing
the oscillation parameters at the current best fit point
$\tan^2\theta=0.44$, $\Delta m^2=6.6\times 10^{-5}$ eV$^2$
\cite{Maltoni:2002ni} we obtain with the Bayesian methods described in
Section~\ref{sec:stat} the 90\CL\ bound
\begin{equation}
\la < 4.0 \times 10^{-10}\mu_B \quad\mbox{(best fit point, solar data)}.
\end{equation}
However, such a bound substantially depends on the values of the neutrino
oscillation parameters. In Fig.~\ref{fig:cnt_rat_sp} we show contours
of the 90\CL\ bound on $\la$ in the $\tan^2\theta - \Delta m^2$ plane.
We find that the bound gets weaker for small values of $\Delta m^2$,
whereas in the upper left part of the LMA-MSW region a bound of the order
$2\times 10^{-10}\mu_B$ is obtained.

By combining solar and reactor data we obtain considerably stronger
bounds. At the best fit point we get at 90\CL
\begin{equation}
\la < 2.0 \times 10^{-10}\mu_B \quad
\mbox{(best fit point, solar + reactor data)}.
\end{equation}
In Fig.~\ref{fig:cnt_sol+reac} we show the contours of the bound in
the $\tan^2\theta - \Delta m^2$ plane for the combination of solar and
reactor data. By comparing Fig.~\ref{fig:cnt_sol+reac} with
Fig.~\ref{fig:cnt_rat_sp} we find that reactor data drastically
improve the bound for low $\Delta m^2$ values. 
\begin{figure}[t] \centering
\includegraphics[width=0.7\linewidth]{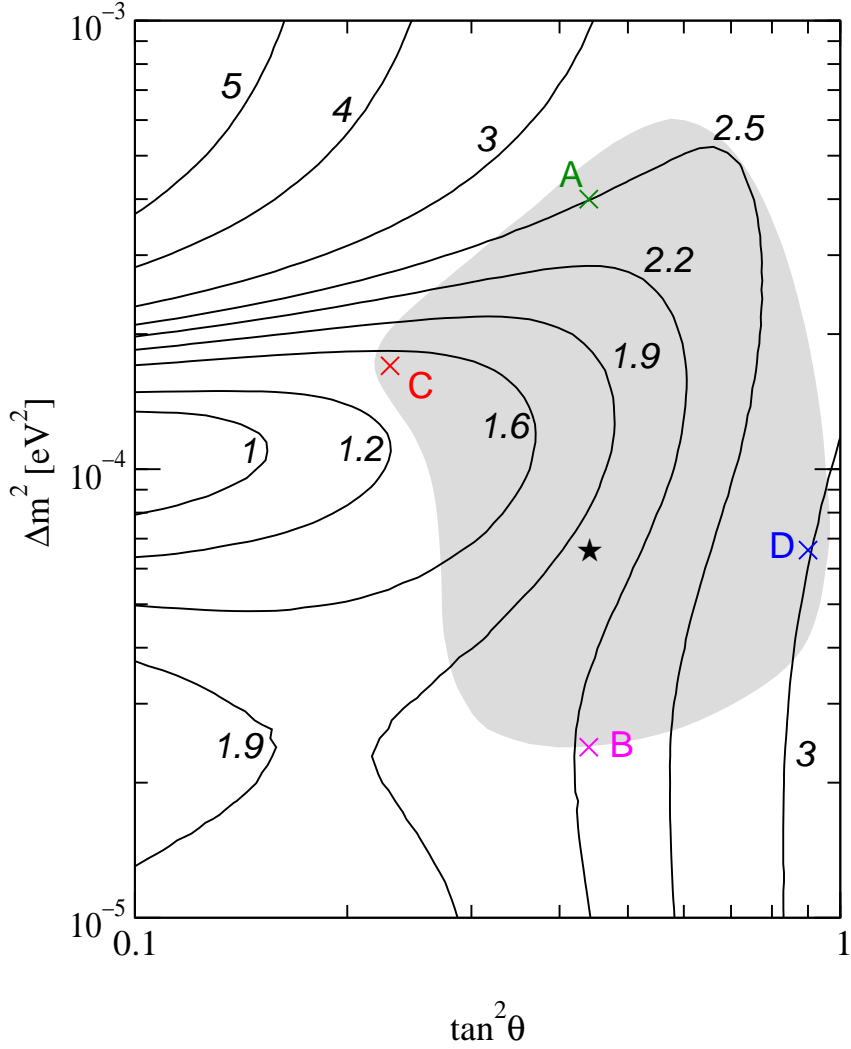}
\caption{Contours of the 90\CL\ bound on $\la$ in units of 
  $10^{-10}\mu_B$ from combined solar and reactor data.  The shaded
  region is the 3$\sigma$ LMA-MSW region from
  Ref.~\cite{Maltoni:2002ni}, the best fit point is marked with a
  star. The crosses labeled A, B, C, D are the test
  points considered in Table \ref{tab:bounds}.}
\label{fig:cnt_sol+reac}
\end{figure}

In order to better understand these results we give in
Tab.~\ref{tab:bounds} bounds on $\la$ for five characteristic points
in the $\tan^2\theta - \Delta m^2$ plane, including the current best
fit point, which is labeled with a \texttt{star}. The other points are
marked in Figs.~\ref{fig:cnt_rat_sp} and \ref{fig:cnt_sol+reac} by
crosses and are labeled A, B, C, D, and the corresponding values of
the oscillation parameters are given in Tab.~\ref{tab:bounds}. Besides
the bound on $\la$ we give also bounds on the individual
$|\Lambda_i|\:(i=1,2,3)$. These bounds are obtained by setting the
other two $\Lambda_i$ to zero.  We show the bounds for the different
data samples: solar rates, the shape of the SK spectrum with free
overall normalization (SK spect), solar rates combined with the SK
spectrum (solar global), reactor data only, solar data combined with
reactor data (sol + react).

Notice that, in contrast with the bound on $\la$ which, thanks to
Eq.~(\ref{basisindep}), is valid in any basis, the bounds on
individual $|\Lambda_i|$ do not have a basis-independent meaning. The
bounds on $|\Lambda_i|$ given in Tab.~\ref{tab:bounds} refer to the
mass-eigenstate 
basis, also used in Ref.~\cite{Schechter:1981hw}.

\begin{table}[!t]\centering
    \catcode`?=\active \def?{\hphantom{0}}
\begin{tabular}{|c|c|c|c|c|c|}
\hline
\rule[-3mm]{0cm}{0.8cm} & solar rates & SK spect & solar global & reactor & sol + react \\
\hline\hline
\rule[-2mm]{0cm}{0.7cm} $\star$ & \multicolumn{5}{|c|}
{$\tan^2\theta=0.44\:,\quad\Delta m^2=6.6\times 10^{-5}$ eV$^2$} \\
\hline
$|\Lambda_1|$ & 1.8 & 1.4 & 1.4 & 2.0 & 1.3 \\
$|\Lambda_2|$ & 5.8 & 4.0 & 4.0 & 1.3 & 1.3 \\
$|\Lambda_3|$ & 1.7 & 1.4 & 1.3 & 1.1 & 1.0 \\ 
\hline
$\la$         & 5.8 & 4.0 & 4.0 & $-$ & 2.0 \\
\hline\hline
\rule[-2mm]{0cm}{0.7cm} A & \multicolumn{5}{|c|}
{$\tan^2\theta=0.44\:,\quad\Delta m^2=4.0\times 10^{-4}$ eV$^2$} \\
\hline
$|\Lambda_1|$ &  3.3 &  2.6 &  2.6 &  2.0 &  2.0 \\
$|\Lambda_2|$ &  3.1 &  2.3 &  2.3 &  1.3 &  1.4 \\
$|\Lambda_3|$ &  2.2 &  1.7 &  1.7 &  1.1 &  1.1 \\
\hline
$\la$         & 3.3 &  2.6 &  2.6 & $-$ &  2.5 \\
\hline\hline
\rule[-2mm]{0cm}{0.7cm} B & \multicolumn{5}{|c|}
{$\tan^2\theta=0.44\:,\quad\Delta m^2=2.4\times 10^{-5}$ eV$^2$} \\
\hline
$|\Lambda_1|$ & ?1.7 & ?1.5 &  ?1.4 & 2.0 &  1.4 \\
$|\Lambda_2|$ & 15.5 & 12.1 & 11.9 &  1.3 &  1.3 \\
$|\Lambda_3|$ & ?1.7 & ?1.5 & ?1.4 &  1.1 &  1.1 \\
\hline
$\la$         & 15.4 & 12.0 &  11.6 & $-$ &  2.3 \\
\hline\hline
\rule[-2mm]{0cm}{0.7cm} C & \multicolumn{5}{|c|}
{$\tan^2\theta=0.23\:,\quad\Delta m^2=1.7\times 10^{-4}$ eV$^2$} \\
\hline
$|\Lambda_1|$ &   2.3 &  1.6 &  1.5 &  2.6 &  1.4\\
$|\Lambda_2|$ &   2.6 &  1.6 &  1.5 &  1.2 &  1.1\\
$|\Lambda_3|$ &   1.7 &  1.1 &  1.1 &  1.1 &  0.9\\
\hline
$\la$         &   2.6 &  1.6 &  1.5 &  $-$ &  1.5\\
\hline\hline
\rule[-2mm]{0cm}{0.7cm} D & \multicolumn{5}{|c|}
{$\tan^2\theta=0.90\:,\quad\Delta m^2=6.6\times 10^{-5}$ eV$^2$} \\
\hline
$|\Lambda_1|$ &   2.2 &  1.9 &  1.9 &  1.6 &  1.6\\
$|\Lambda_2|$ &   8.3 &  6.3 &  6.4 &  1.5 &  1.6\\
$|\Lambda_3|$ &   2.2 &  1.9 &  1.8 &  1.1 &  1.2\\
\hline
$\la$         &   8.2 &  6.2 &  6.2 & $-$ & 3.0\\
\hline
\end{tabular}
\caption{90\% C.L. bounds on $|\Lambda_i|\:(i=1,2,3)$ and on $\la$ in
units of $10^{-10}\mu_B$ at fixed points in the $\tan^2\theta - \Delta
m^2$ plane from different data samples. The hyphen 
indicates that $\la$ cannot be bounded from reactor data alone.}
\label{tab:bounds}
\end{table}

By comparing the first and the second column of Tab.~\ref{tab:bounds}
we notice that both solar rates and SK spectrum are important:
although the bound from the SK spectrum is slightly stronger, the one
which follows from solar rates is in general of the same order.  The
bound we obtain by combining rates and SK spectrum is displayed in the
third column.
Concerning the reactor data, clearly these bounds do not depend
on $\Delta m^2$, and the bound on $|\Lambda_3|$ is even 
completely independent of
the neutrino oscillation parameters (see Eq.~(\ref{finalreactor})).
As mentioned above, reactor data on their own cannot be used to 
constrain $\la$.

In the upper parts of the LMA-MSW region, the solar data alone give already
a strong bound on $\la$. From the bounds given in
Tab.~\ref{tab:bounds} at point A we see that for such high values of
$\Delta m^2$ all $|\Lambda_i|$ are strongly constrained by solar data.
This is due to the fact that there the probability $P_{e1}$ relevant in SK
is close to the optimal value of 0.5 (see Fig.~\ref{fig:Pe1}). In this
parameter region the combination with reactor data does not improve
the bound significantly. This is also true for the region of small
$\tan^2\theta$ around point C.

The behavior of the bounds for low $\Delta m^2$ values can be
understood by looking at point B. In this $\Delta m^2$ region the
probability $P_{e1}$ is very small for the neutrino energies relevant
for SK, as seen in Fig.~\ref{fig:Pe1}. Therefore, solar data give
only a very weak constraint on $|\Lambda_2|$ as can be seen from
Tab.~\ref{tab:bounds}, and the bound on $\la$ is dominated by this
weak bound on $|\Lambda_2|$ (see Eq.~(\ref{mumin2})) leading to the
rather poor bound of $11.6 \times 10^{-10}\mu_B$ from solar data. In
this case the combination with reactor data improves the bound
drastically to $2.3 \times 10^{-10}\mu_B$. 

Up to now we have calculated bounds on neutrino \TMs\ for fixed values
of the oscillation parameters $\tan^2\theta$ and $\Delta m^2$ (method
I described in Section~\ref{sec:stat}). These results will be especially
useful after KamLAND will have determined the oscillation parameters
with good accuracy. In the following we change our strategy and
minimize the $\chi^2$ for each value of $\la$ with respect to
$\tan^2\theta$ and $\Delta m^2$ (method II). This will lead to a bound
on $\la$ taking into account the current knowledge concerning the
oscillation parameters.
To this end we make use of the global solar neutrino data (including
earth matter effects) as described in Ref.~\cite{Maltoni:2002ni} in
order to obtain the correct $\chi^2$ behavior as a function of
$\tan^2\theta$ and $\Delta m^2$. Only in the expression for
$\mu^2_\mathrm{LMA}$ we neglect earth matter effects as described in
Section~\ref{sec:theor-fram}, since they are known to be small in the
LMA-MSW region~\cite{earthmatter}. Performing this analysis we obtain
the following bounds at 90\CL:
\begin{equation}\label{unconstrbounds}
\la < \left\{ \begin{array}{l@{\quad}l}
6.3 \times 10^{-10}\mu_B  & \mbox{(unconstrained, solar data)} \\
2.0 \times 10^{-10}\mu_B  &
\mbox{(unconstrained, solar + reactor data).}
\end{array}\right.
\end{equation}

Together with Figs.~\ref{fig:cnt_rat_sp} and \ref{fig:cnt_sol+reac}
the bounds given in Eq.~(\ref{unconstrbounds}) constitute an important
result of this paper. They show that, assuming the solar LMA-MSW
solution, current solar neutrino data can be used to constrain
\texttt{all} elements of the Majorana neutrino \TM\
matrix. A combination with data from reactor experiments significantly
strengthens the bound on $\la$.

\section{Simulation of the Borexino experiment}
\label{sec:sensitivity-borexino}

Here we investigate the sensitivity of the Borexino
experiment~\cite{borexino,Alimonti:2002xc} to neutrino \TMs. This
experiment is mainly sensitive to the solar $^7$Be neutrino flux,
which will be measured by elastic neutrino--electron scattering.
Therefore, Borexino is similar to SK, the main difference is the
mono-energetic line of the $^7$Be neutrinos, with an energy of 0.862
MeV, which is roughly one order of magnitude smaller than the energies
of the $^8$B neutrino flux relevant in SK.

\subsection{The Borexino $\chi^2$-function}
\label{sec:borexinochi2}

To estimate the sensitivity of Borexino we consider the following 
$\chi^2$-function:
\begin{equation}
\chi^2_\mathrm{borexino} =
\sum_{i,j=1}^{N_\mathrm{bins}} 
(N_i^\mathrm{th} - D_i) (V^\mathrm{borex}_{ij})^{-1} (N_j^\mathrm{th} - D_j) \,.
\end{equation}
Here $N_i^\mathrm{th}$ is the theoretical prediction for the number of
events in the electron recoil energy bin $i$, and $D_i$ is the
(hypothetical) observed number of events. Borexino will observe recoil
electrons with kinetic energy in the range 0.25 to 0.8 MeV. As a
realistic example we adopt $N_\mathrm{bins}=8$ bins in electron
recoil energy and an energy resolution of 0.058 
MeV~\cite{borexino,Alimonti:2002xc,Berezhiani:2001rt}. We have
checked that our results are essentially independent of the exact
value of the energy resolution.

The theoretical prediction for the number of events in the bin $i$ after
$N_\mathrm{years}$ years of Borexino data taking is given by
\begin{equation}\label{Nth}
N_i^\mathrm{th} = N_i^\mathrm{weak} +
N_i^\mathrm{em}(\mu_\mathrm{LMA}) + 
N_i^\mathrm{bg} 
\,, \quad N_i^x = 365 \, N_\mathrm{years} \, n_i^x
\end{equation}
with $x=$ th, weak, em, bg, and $n_i^\mathrm{weak,em,bg}$ is the
number of events per day from weak scattering, electromagnetic
scattering and the background, respectively. If not stated otherwise
we consider a running time of $N_\mathrm{years}=3$ years. In order to
estimate the sensitivity of Borexino for neutrino \TMs\ we assume that
the data are generated by neutrinos \texttt{without} \TMs:
\begin{equation}
D_i = N_i^\mathrm{weak} + N_i^\mathrm{bg} \,.
\end{equation}
Hence, we obtain
\begin{equation}\label{chi2bor}
\chi^2_\mathrm{borexino} =
\sum_{i,j=1}^{N_\mathrm{bins}} 
N_i^\mathrm{em}\,  (V^\mathrm{borex}_{ij})^{-1} \, N_j^\mathrm{em} \,.
\end{equation}
The minimum of this $\chi^2$ occurs for $\mu_\mathrm{LMA}=0$ and is
always zero. With the $\chi^2$ of Eq.~(\ref{chi2bor}) we calculate a
bound on $\la$ at a given C.L. This bound corresponds to the maximum
allowed value of $\la$ which cannot be distinguished from $\la = 0$,
and is therefore a measure for the obtainable sensitivity to $\la$ at
Borexino.

The numbers of events due to the weak and electromagnetic scattering are
calculated by
\begin{eqnarray}
n_i^\mathrm{weak} &=& \mathcal{N}
\sum_f  \Phi_f^0 \int_0^\infty dE_\nu\, \phi_f
\left\{ P_{ee}^f \, [\sigma^W_e]_i +
(1- P_{ee}^f) [\sigma^W_\mu ]_i \right\}, \nonumber\\
n_i^\mathrm{em}&=& \mathcal{N}
\sum_f  \Phi_f^0 \int_0^\infty dE_\nu\, \phi_f \,
\mu^2_\mathrm{LMA} \, [\sigma'_\mathrm{em}]_i \,, \label{borrates}
\end{eqnarray}
respectively.
The effective cross sections in the bin $i$ are calculated like in
Eqs.~(\ref{sigma}) and (\ref{sigmaem}), by choosing the lower and upper
bounds of the recoil energy interval $T_1$ and $T_2$ according to the
given bin $i$.
We sum over the four most important neutrino fluxes ($f$ = Be, pep, N, O)
in Borexino (see Ref.~\cite{Alimonti:2002xc}). For the mono-energetic
lines we have $\phi_\mathrm{Be}(E_\nu)=\delta(E_\nu - 0.862
\,\mathrm{MeV})$ and $\phi_\mathrm{pep}(E_\nu)=\delta(E_\nu - 1.442
\,\mathrm{MeV})$, and we use the normalized spectra
$\phi_\mathrm{N}(E_\nu)$ and $\phi_\mathrm{O}(E_\nu)$ given in
Ref.~\cite{bahcallpage}. The absolute values of the neutrino fluxes
$\Phi_f^0$ from Ref.~\cite{BP00} are given by
$\Phi^0_\mathrm{Be}=0.429 ,\, \Phi^0_\mathrm{pep}= 0.014 ,\,
\Phi^0_\mathrm{N}=0.055 ,\, \Phi^0_\mathrm{O}=0.048$ in units of
$10^{10}$ cm$^{-2}$s$^{-1}$.  Note that from the two $^7$Be lines the
line at 0.38 MeV, which constitutes 10\% of the total $^7$Be flux,
does not contribute to the signal in Borexino.  Therefore the value
for $\Phi^0_\mathrm{Be}$ given above has been obtained by multiplying
the value given in Ref.~\cite{BP00} with 0.9.

\begin{table}
\begin{center}
\begin{tabular}{c|ccccc}
\hline\hline
flux & $^7$Be & pep & $^{13}$N & $^{15}$O & background\\
\hline
events/day & 43.3 & 2.0 & 4.0 & 5.5 & 19\\
\hline\hline
\end{tabular}
\end{center}
\caption{Expected number of events per day
in Borexino for the kinetic energy of the recoil electrons in the
range 0.25 to 0.8 MeV resulting from different components of the 
SSM solar neutrino flux \texttt{without} neutrino conversion (taken from
Table 4 of Ref.~\cite{Alimonti:2002xc}). Also shown is the expected number of
background events per day.}\label{borexinotable}
\end{table}

In Table \ref{borexinotable} we show the expected number of events per
day in Borexino for the kinetic energy of the recoil electrons in the
range 0.25 to 0.8 MeV, resulting from different solar neutrino fluxes,
for the standard solar model without neutrino oscillations.  We do not
consider other neutrino fluxes, because they contribute to the signal
with less than 0.2 events per day. We fix the normalization constant
$\mathcal{N}$ in Eq.~(\ref{borrates}) in such a way that the number of
events per day from the $^7$Be line for $P_{ee}=1$ and
$\mu_\mathrm{LMA}=0$ is 43.3. Then we can reproduce the number of events
arising from the other fluxes as given in Table \ref{borexinotable}
to good accuracy.

The number of background events $n_i^\mathrm{bg}$ is obtained in the
following way. We read off the shape of the background as a function of
the electron energy from Fig.~15 of Ref.~\cite{Alimonti:2002xc} and
normalize this function to 19 background events per day in the energy
range 0.25 to 0.8 MeV~\cite{borexino}. Then we
can calculate $n_i^\mathrm{bg}$ by integrating over the corresponding
energy interval for each bin. 

We use the following covariance matrix in Eq.~(\ref{chi2bor}):
\begin{equation}
V^\mathrm{borex}_{ij} = \delta_{ij} N_i^\mathrm{th} 
+ 
\sum_{f_1,f_2} N_{i f_1} N_{j f_2}
\sum_{\beta=1}^{12} \alpha_{f_1 \beta}\alpha_{f_2 \beta}
\left( \Delta\!\ln\!X_\beta \right)^2
+
N_i^\mathrm{bg} N_j^\mathrm{bg} (\Delta \ln N^\mathrm{bg})^2 \,.
\end{equation}
The first term is the statistical uncertainty, taken as the
square-root of the predicted number of events. The second term
describes the uncertainty in the solar neutrino fluxes. Here $N_{if}$
are the contributions of the individual solar neutrino fluxes to the
total event numbers in each bin: $N_i^\mathrm{th}= \sum_f N_{if} +
N_i^\mathrm{bg}$, and the sum is over the four fluxes relevant in
Borexino (for definition and values of the quantities $\alpha_{f\beta}$ and
$X_\beta$ see Ref.~\cite{fogli}). The last term takes into account a
systematic uncertainty in the number of background events. We assume
full correlation between the bins and that the relative uncertainty is
the same for all bins: $\Delta N_i^\mathrm{bg}/N_i^\mathrm{bg} =
\Delta \ln N^\mathrm{bg}$. We adopt a standard value for 
our calculations of $\Delta \ln N^\mathrm{bg}=10\%$.

\begin{figure}[!t] \centering
\includegraphics[width=0.65\linewidth]{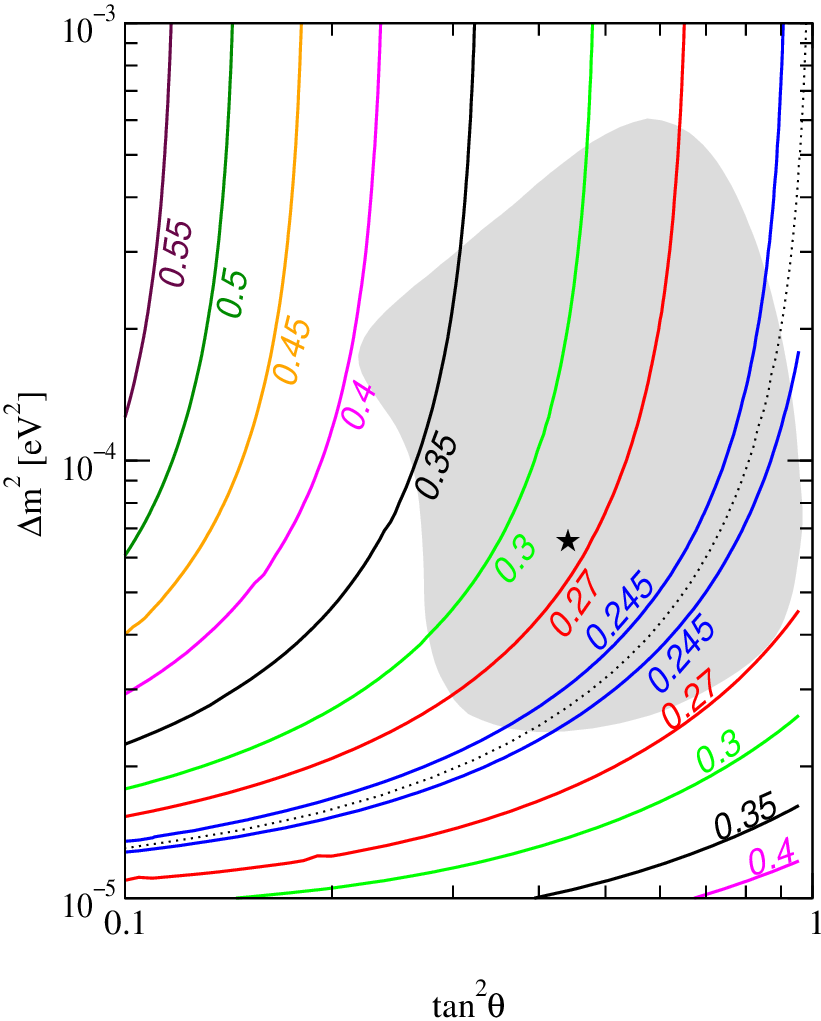}
\caption{Contours of the 90\CL\ bound on $\la$ after 3 years of
  Borexino data-taking in units of $10^{-10}\mu_B$. The current best
  fit point is shown by the star, and the shaded region is the allowed
  LMA-MSW region at 3$\sigma$, from Ref.~\cite{Maltoni:2002ni}. The
  dotted line corresponds to $P_{e1}=0.5$ for $^7$Be neutrinos.}
\label{fig:borexContour}
\end{figure}

\subsection{Sensitivity on $\la$ of Borexino}

Using the statistical method described in Sections \ref{sec:stat} and
\ref{sec:borexinochi2} we obtain for the current best fit point
$\Delta m^2 = 6.6\times10^{-5}$ eV$^2$, $\tan^2\theta = 0.44$ the upper
bound (sensitivity)
\begin{equation}\label{borexinobound}
\la \le 0.28 \times 10^{-10} \mu_B \quad\mbox{at}\quad 90\%\:\mbox{C.L.}
\end{equation}
after three years of Borexino data taking. This bound is about one
order of magnitude stronger than the bound from existing data. We have
checked that a combined analysis of Borexino with existing data (solar
as well as reactor data) does not improve the bound of
Eq.~(\ref{borexinobound}). The bound depends on the actual value of
the oscillation parameters; in Fig.~\ref{fig:borexContour} we show
contours of the 90\CL\ bound in the $\tan^2\theta-\Delta m^2$ plane.
The strongest attainable limit is roughly $0.24\times 10^{-10} \mu_B$.
In agreement with the discussion we gave in connection with
Eq.~(\ref{mumin2}) we find that the strongest bound is obtained when
$P_{e1}=0.5$, as shown by the dotted line in
Fig.~\ref{fig:borexContour}.

\begin{figure}[t] \centering
\includegraphics[width=0.8\linewidth]{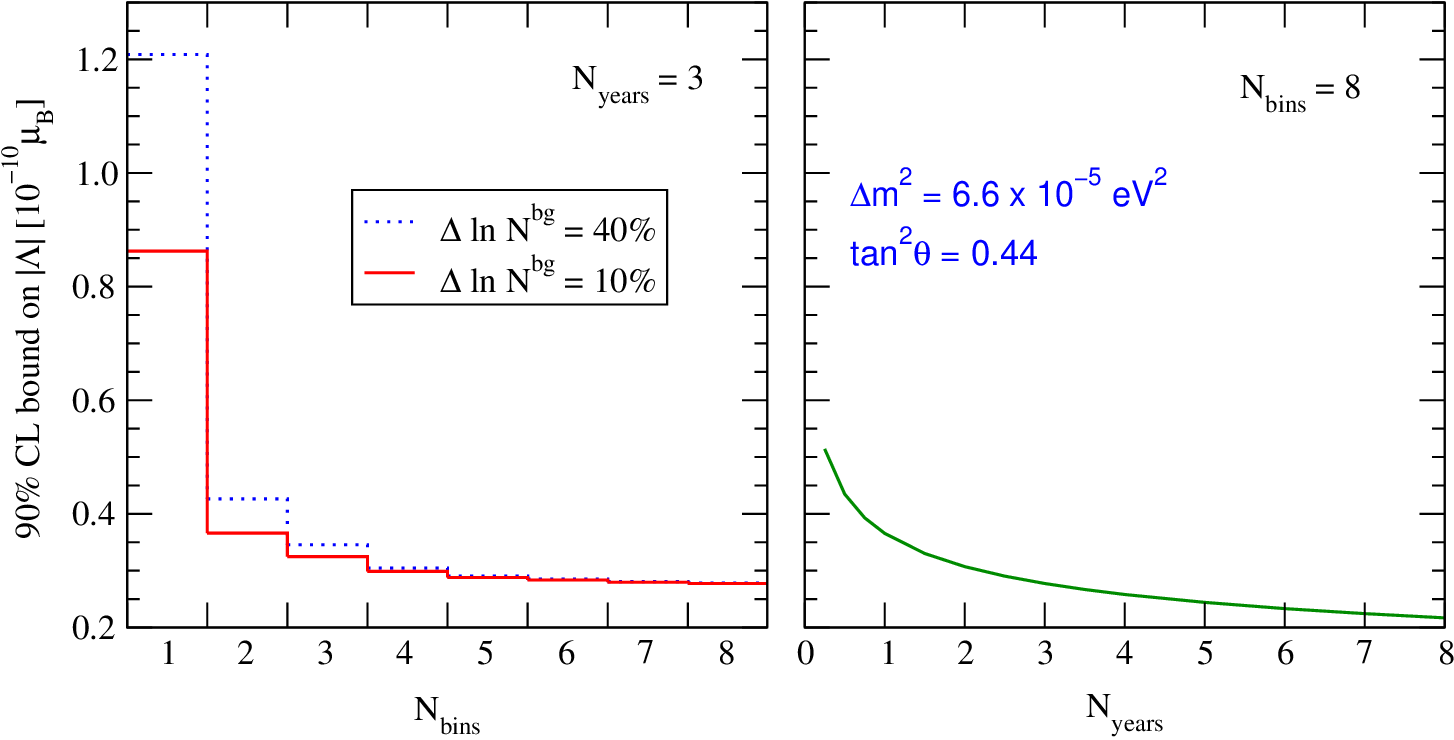}
\caption{The left panel shows the 90\CL\ bound on $\la$ after 3 years of
  Borexino data-taking as a function of the number of recoil electron
  energy bins for two values of the error on the background. The right
  panel gives the 90\CL\ bound on $\la$ as a function of Borexino
  running time for  $\Delta\ln N^\mathrm{bg}=10$\%.}
\label{fig:bortest}
\end{figure}
We have performed several tests concerning our assumptions for
Borexino. Fig.~\ref{fig:bortest} (left panel) shows the dependence of
the bound on the number of bins in the recoil electron energy. We can see
from this figure that a measurement of the total rate alone
($N_\mathrm{bins}=1$) gives a rather weak bound. However a few bins
are already enough for a considerable improvement of the bound on
$\la$.  It is interesting to see that the error on the background is
only important for few bins, and becomes completely unimportant for
more than 3 or 4 bins, because the main information comes from the
spectral shape.

In the right panel in Fig.~\ref{fig:bortest} we show the bound as a
function of the Borexino running time. Here we adopt again our nominal
value of 8 energy bins. After one year a sensitivity of $0.4\times
10^{-10} \mu_B$ can be obtained with Borexino, which is already about
one order of magnitude stronger than the current bound. After three years
of data-taking only a minor improvement of the bound is possible,
mainly due to the uncertainty in the solar neutrino
fluxes.\footnote{Note that we do not include any experimental
  systematic uncertainties in our Borexino analysis.}

\section{Conclusions}
\label{sec:conclusions}

In this paper we have presented stringent bounds on electromagnetic
Majorana transition moments (\TMs). Such \TMs---if present---would
contribute to the elastic neutrino--electron scattering signal in
solar neutrino experiments like Super-Kamiokande or the upcoming
Borexino experiment, as well as in reactor neutrino experiments.
Motivated by the robust status of the LMA-MSW solution of the solar
neutrino problem indicated by recent data, and also by the prospects
of its possible confirmation by the KamLAND experiment, we have taken
this solution as basis for our investigation. In this context we have
clarified which effective magnetic moments are probed in solar and
reactor experiments---see Eqs.~(\ref{finalLMA}) and (\ref{reactorMM}),
respectively. Using most recent global solar neutrino data we have
performed a fit in terms of the oscillation parameters and the
elements of the complex \TM\ matrix $\lambda$ of three active Majorana
neutrinos. Taking into account the antisymmetry of the \TM\ matrix by
the parameterization $\lambda_{jk} = \varepsilon_{jkl}\Lambda_l$, we have
shown that solar neutrino data allow to constrain $\la$,
\textit{i.e.}, \texttt{all} elements of the \TM\ matrix are bounded at
the same time. We want to stress that the bounds on $\la$ hold in any
basis, because $\la$ is an intrinsic neutrino property; in terms of
\TMs\ it is given by
\begin{equation}
\la^2 =
|\lambda_{e\mu}|^2 + |\lambda_{\mu\tau}|^2 + |\lambda_{\tau e}|^2 =
|\lambda_{12}|^2 + |\lambda_{23}|^2 + |\lambda_{31}|^2
\end{equation}
in the flavour and mass-eigenstate bases, respectively.
On the other hand, the
bounds on the individual $|\Lambda_i|$ ($i=1,2,3$) in
Tab.~\ref{tab:bounds} refer only to 
the mass-eigenstate basis.

A fit to the global solar neutrino data leads to the bound $\la <
6.3\times 10^{-10}\mu_B$ at 90\CL\ We have also considered the role of
reactor neutrino data on neutrino \TMs, shown to be complementary to
solar neutrino data.  A combined fit of reactor and solar data leads
to significantly improved bounds: at 90\CL\ we get $\la < 2.0\times
10^{-10}\mu_B$. In the very near future the KamLAND experiment will
crucially test the LMA-MSW solution. If KamLAND confirms the LMA-MSW
solution then a precise determination of the parameters $\tan^2\theta$
and $\Delta m^2$ might be possible. This motivated us to scan the
$\tan^2\theta - \Delta m^2$ plane and to calculate the corresponding
bound on $\la$ in each point.  These results are shown in
Figs.~\ref{fig:cnt_rat_sp} and \ref{fig:cnt_sol+reac} for solar data
and solar + reactor data, respectively.

We have also investigated the potential of the upcoming
neutrino--electron scattering solar neutrino experiment Borexino to
constrain neutrino \TMs.  Performing a detailed simulation of the
experiment we find that it will improve the bound on $\la$ by about
one order of magnitude with respect to present bounds.

Last, but not least, let us mention that alternative solutions
to the solar neutrino problem based on non-standard neutrino matter
interactions~\cite{Guzzo:2001mi}, which may arise in models of
neutrino masses~\cite{NSImodels}, can also be constrained in an
analogous way~\cite{Berezhiani:2001rt}. If, as seems likely, the
LMA-MSW solution is finally borne out, the improved determination of
the oscillation parameters expected, say at KamLAND, will provide an
extremely useful stepping stone for probing other features of neutrino
physics beyond the Standard Model.

\section*{Acknowledgments}

This work was supported by Spanish grant PB98-0693, by
the European Commission RTN network HPRN-CT-2000-00148 and by the
European Science Foundation network grant N.~86. T.S.~has been
supported by the DOC fellowship of the Austrian Academy of Science
and, in the early stages of this work, by a fellowship of the European
Commission Research Training Site contract HPMT-2000-00124 of the host
group.  M.M.~was supported by contract HPMF-CT-2000-01008 and M.A.T. 
was supported by the M.E.C.D.\ fellowship AP2000-1953.

\end{document}